\documentclass[11pt]{article}
\usepackage{amssymb,amsmath,amsthm,amscd,latexsym}
\usepackage{mathrsfs}
\usepackage{mathrsfs}
\usepackage{amsfonts}
\usepackage{amsmath}
\usepackage{amssymb}
\usepackage{amscd}
\usepackage{enumitem}
\usepackage[all]{xy}
\usepackage{CJK}
\usepackage{xypic}
\usepackage{graphicx}
\usepackage{epstopdf}

\renewcommand{\paragraph}{\roman{paragraph}}
\input cyracc.def

\begin{document}
\begin{CJK*}{GBK}{song}\CJKtilde
\title{\bf Skew cyclic codes over $\mathbb{F}_{q}+v\mathbb{F}_{q}+v^{2}\mathbb{F}_{q}$
\thanks{\textbf{Foundation item:} Supported by NNSF of China (61202068),
Talents youth Fund of Anhui Province Universities (2012SQRL020ZD).
\textbf{Biography:}SHI Min-jia (corresponding author), male, born in 1980, PhD. Research field: coding theory and cryptography. E-mail: smjwcl.good @163. com. \textbf{This manuscript was finished on 2014/12/01, now it was accepted for publication by a magazine, which was submitted on 2015/02/03.}}}
\author{\small{Minjia SHI, Ting YAO}\\ \small{(School of Mathematical Sciences of Anhui University. Anhui, China.)}\\
\small{Adel Alahmadi, Patrick Sol\'e}\\ \small{(Department of Mathematics of King Abdulaziz University. Jeddah, Saudi Arabia)}}
\date{}
\maketitle
\end{CJK*}

\begin{CJK}{GBK}{song}
{\bf Abstract:} {\normalsize   In this article, we study skew cyclic codes over ring $R=\mathbb{F}_{q}+v\mathbb{F}_{q}+v^{2}\mathbb{F}_{q}$, where $q=p^{m}$, $p$ is an odd prime and $v^{3}=v$. We describe generator polynomials of skew cyclic codes over this ring and investigate the structural properties of skew cyclic codes over $R$ by a decomposition theorem. We also describe the generator polynomials of the duals of skew cyclic codes. Moreover, the idempotent generators of skew cyclic codes over $\mathbb{F}_{q}$ and $R$ are considered.}


\section{Introduction}

\hspace*{0.6cm}Codes over finite rings have been studied since the early 1970s, because of their rich structure, linear codes are the most frequent in coding theory. While different approaches have been applied to produce certain types of codes with good parameters and properties. In [8], Hammons et al. showed that some important binary nonlinear codes can be obtained from cyclic codes over $\mathbb{Z}_{4}$ through the Gray map. Recently, in [2], D. Boucher et al. introduced the class of $\theta$-cyclic (skew cyclic) codes that generalizes the concept of cyclic codes over non-commutative polynomial rings, called a skew polynomial ring, to construct these types of codes.

In[2], D. Boucher et al. gave skew cyclic codes defined by using the skew polynomial ring with an automorphism $\theta$ over the finite field with $q$ elements. The polynomial ring is denoted by $\mathbb{F}_{q}[x,\theta]$, where the addition is the usual polynomial addition and the multiplication is defined by the rule $xa=\theta(a)x,(a\in \mathbb{F}_{q})$ in [9] which means the finite field elements are not commutative with the indeterminate $x$. In [3], D. Boucher and F. Ulmer showed that the dual of a $\theta$-cyclic code is still a $\theta$-cyclic code. I. Siap et al. [10] gave the structure of skew cyclic codes of arbitrary length. J. Gao in [6] studied skew cyclic codes over $\mathbb{F}_{p}+v\mathbb{F}_{p}$. In [7], F. Gursoy et al. presented the construction of skew cyclic codes over $\mathbb{F}_{q}+v\mathbb{F}_{q}$ for different automorphisms. Moreover, T. Abualrub et al. [1] and D. Boucher et al. [4] studied skew quasi-cyclic codes and skew constacyclic codes, respectively.

In this paper, we study skew cyclic codes defined by the skew polynomial ring with coefficients over ring $R=\mathbb{F}_{q}+v\mathbb{F}_{q}+v^{2}\mathbb{F}_{q}$, where $q=p^{m}, \ p$ is an odd prime and $v^{3}=v$. In our work, we consider the automorphisms $$\theta_{i}: \mathbb{F}_{q}+v\mathbb{F}_{q}+v^{2}\mathbb{F}_{q}\rightarrow \mathbb{F}_{q}+v\mathbb{F}_{q}+v^{2}\mathbb{F}_{q}$$
$$a+bv+cv^{2}\mapsto a^{p^{i}}+vb^{p^{i}}+v^{2}c^{p^{i}}.$$
Denote the skew polynomial ring as $R[x,\theta_{i}]$, where the addition is the usual polynomial addition and the multiplication is defined by the rule $xa=\theta_{i}(a)x, (a\in R)$.

The rest of the paper is organized as follows: Section 2 gives a Gray map from $R$ to $\mathbb{F}_{q}^{3}$. In Section 3, we mainly describe the basic properties of linear codes over $R$ and their structures. In Section 4, we describe the generator polynomials of skew cyclic codes and the duals of skew cyclic codes. We prove that every skew cyclic code over $R$ is principally generated and give the idempotent generators of $\mathbb{F}_{q}$ and $\mathbb{F}_{q}+v\mathbb{F}_{q}+v^{2}\mathbb{F}_{q}$.
\section{Preliminary}

\hspace*{0.6cm}Let $R=\mathbb{F}_{q}+v\mathbb{F}_{q}+v^{2}\mathbb{F}_{q}$, where $q=p^{m}$, $p$ is an odd prime and $v^{3}=v$. Clearly, $R\cong \mathbb{F}_{q}[v]/(v^{3}-v)$. $R$ is a commutative ring with identity and characteristic $p$. For any element $r$ of $R$ , $r$ can be expressed uniquely as $r=a+bv+cv^{2}$, where $a,b,c\in \mathbb{F}_{q}$. It is easily checked that $R$ is a Frobenius ring but not local. $R$ is also principal and has three maximal ideals $\langle v\rangle, \langle v-1\rangle$ and $\langle v+1\rangle$.

From [5], we have the following definition.

\textbf{Definition 2.1} The definition of the Gray map on $R^{n}$ as follows
$$\Phi: R^{n} \rightarrow \mathbb{F}^{3n}_{q}$$
$$(r_{0},r_{1},\ldots,r_{n-1})\rightarrow (a_{0},a_{0}+b_{0}+c_{0},a_{0}-b_{0}+c_{0},\ldots,a_{n-1},a_{n-1}+b_{n-1}+c_{n-1},a_{n-1}-b_{n-1}+c_{n-1}),$$
where $r_{i}=a_{i}+b_{i}v+c_{i}v^{2}, i=0,1,\ldots,n-1$.\

\textbf{Definition 2.2} Let $r=a+bv+cv^{2}$ be an element of $R$, then the Lee weight of $r$ is defined as $$\omega_{L}(r)=\omega_{H}(a,a+b+c,a-b+c),$$
where the symbol $\omega_{H}(v)$ denotes the Hamming weight of $v$ over $\mathbb{F}_{q}$.
\section{Linear codes over $R$ }

\hspace*{0.6cm}In this section, we generalize the structure and properties from [5] to codes over $R$. Hence the proofs of many of the theorems will be omitted.

\textbf{Lemma 3.1} ([5, Lemma 1]) The Gray map $\Phi$ is a distance-preserving map from $(R^{n}, \ {\rm Lee \ distance})$ to $(\mathbb{F}^{3n}_{q}, \ {\rm Hamming \ distance})$ and it is also $\mathbb{F}_{q}$-linear.

According to the definition of the Gray map $\Phi$ and Lemma 3.1, we have the following lemma.

\textbf{Lemma 3.2} Let $C$ be a linear code of length $n$ over $R$ with rank $k$ and minimum Lee distance $d$, then $\Phi(C)$ is a $[3n, k, d]$ linear code over $\mathbb{F}_{q}$.

\textbf{Proof}
From Lemma 3.1, we see that $\Phi(C)$ is a $\mathbb{F}_{q}$-linear code. From the definition of the Gray map. We can easily obtain that $\Phi(C)$ has dimension $k$ and length $3n$ since $\Phi$ is a bijective map from $R^{n}$ to $\mathbb{F}^{3n}_{q}$. Moreover, since Gray map $\Phi$ is a distance-preserving map, so $\Phi(C)$ has the same minimum distance $d$.

Let $C$ be a linear code over $R$. The dual of $C$ consists of all vectors of $R^{n}$ which are orthogonal to every codeword in $C$. A code $C$ is said to be self-dual (resp. self-orthogonal) if $C=C^{\bot}$ (resp. $C\subseteq C^{\bot}$). Now, in light of Ref.[5], we give the following theorem.

\textbf{Theorem 3.1} ([5, Theorem 1]) Let $C$ be a linear code over $R$, then $\Phi(C)^{\bot}=\Phi(C^{\bot})$. Moreover, if $C$ is self-dual, so is $\Phi(C)$ over $\mathbb{F}_{q}$.

By the Chinese Remainder Theorem, we have
\begin{eqnarray*}
R&=&(1-v^{2})R\oplus(2^{-1}v+2^{-1}v^{2})R\oplus(-2^{-1}v+2^{-1}v^{2})R\\
 &=&(1-v^{2})\mathbb{F}_{q}\oplus(2^{-1}v+2^{-1}v^{2})\mathbb{F}_{q}\oplus(-2^{-1}v+2^{-1}v^{2})\mathbb{F}_{q}.
\end{eqnarray*}

For the sake of convenience, we denote by $\eta_{1}, \eta_{2}, \eta_{3}$ respectively the following elements of $R$.
$$\eta_{1}=1-v^{2}, \eta_{2}=2^{-1}v+2^{-1}v^{2}, \eta_{3}=-2^{-1}v+2^{-1}v^{2}.$$
Note that $\eta_{1}, \eta_{2},$ and $\eta_{3}$ are mutually orthogonal idempotents over $R$ and $\eta_{1}+\eta_{2}+\eta_{3}=1.$

Let $C$ be a linear code of length $n$ over $R$. Define
\begin{eqnarray*}
C_{1}=\{x\in \mathbb{F}_{q}^{n}|\exists y,z\in \mathbb{F}_{q}^{n},\eta_{1}x+\eta_{2}y+\eta_{3}z\in C\},\\
C_{2}=\{y\in \mathbb{F}_{q}^{n}|\exists x,z\in \mathbb{F}_{q}^{n},\eta_{1}x+\eta_{2}y+\eta_{3}z\in C\},\\
C_{3}=\{z\in \mathbb{F}_{q}^{n}|\exists x,y\in \mathbb{F}_{q}^{n},\eta_{1}x+\eta_{2}y+\eta_{3}z\in C\}.
\end{eqnarray*}
Then $C_{1},C_{2},C_{3}$ are all linear codes of length $n$ over $\mathbb{F}_{q}$. Moreover, the code $C$ of length $n$ over $R$ can be uniquely expressed as
\begin{eqnarray*}
C=\eta_{1}C_{1}\oplus\eta_{2}C_{2}\oplus\eta_{3}C_{3}.
\end{eqnarray*}

Let $G_{1}, G_{2}$ and $G_{3}$ be the generator matrices of $C_{1}, C_{2}$ and $C_{3}$, respectively, then
\begin{equation}
\label{generator-C1} G=\left(
\begin{array}{cccccc}
\eta_{1}G_{1} \\
\eta_{2}G_{2} \\
\eta_{3}G_{3}
\end{array} \right)\nonumber
\end{equation}
is the generator matrix of $C$.

According to Definition 2.1, we can easily obtain the following proposition.

\textbf{Proposition 3.1} If $C$ is a linear code of length $n$ over $R$ with generator matrice $G$, then we have
\begin{equation}
\label{generator-C1} \Phi(G)=\left(
\begin{array}{cccccc}
\Phi(\eta_{1}G_{1}) \\
\Phi(\eta_{2}G_{2}) \\
\Phi(\eta_{3}G_{3})
\end{array} \right)=\left(\begin{array}{cccccc}
G_{1} & 0 & 0 \\
0 & G_{2} & 0 \\
0 & 0 & G_{3}
\end{array}\right).\nonumber
\end{equation}
Moreover, $d_{H}(\Phi(C))=min\{d_{H}(C_{1}), d_{H}(C_{2}), d_{H}(C_{3})\}$.

\textbf{Theorem 3.2} ([5, Theorem 3]) Let $C$ be a linear code of length $n$ over $R$, then
$$C^{\perp}=\eta_{1}C_{1}^{\perp}\oplus\eta_{2}C_{2}^{\perp}\oplus\eta_{3}C_{3}^{\perp}.$$
Moreover, $C$ is self-dual if and only if $C_{1}, C_{2}$ and $C_{3}$ are all self-dual over $\mathbb{F}_{q}$.
\section{Skew cyclic codes over $R$}

\hspace*{0.6cm}In this section, we mainly study skew cyclic codes over $R$ with automorphism $\theta_{i}$ and give the generator polynomials of skew cyclic codes and their dual codes. Let us denote the order of $\theta_{i}$, which is $t_{i}=\frac{m}{i}$ for some positive integer. In the commutative case if $(n,q)=1$, then every cyclic code of length $n$ over $\mathbb{F}_{q}$ has a unique idempotent generator. Note that in the skew polynomial ring $\mathbb{F}_{q}[x,\theta_{i}]$, if $(n, t_{i})=1$, then the factorization of $x^{n}-1$ in $\mathbb{F}_{q}[x,\theta_{i}]$ is unique (see [7]). In this part, we also show that a formula for the number of skew cyclic codes of length $n$ over $R$ when $(n,t_{i})=1$.

We first give the concept of skew cyclic codes over $R$.

\textbf{Definition 4.1} Let $R$ be a ring and $\theta_{i}$ be an automorphism of $R$. A linear code $C$ of length $n$ over $R$ is a skew cyclic code with the property that
\begin{equation*}
c=(c_{0},c_{1},\ldots,c_{n-1})\in C\Rightarrow\sigma(c)=(\theta_{i}(c_{n-1}),\theta_{i}(c_{0}),\ldots,\theta_{i}(c_{n-2}))\in C,
\end {equation*}
where $\sigma(c)$ is a skew cyclic shift of $c$.

In polynomial representation, the codewords $(c_{0},c_{1},\ldots,c_{n-1})$ of a skew cyclic code are coefficient tuples of elements $c_{n-1}x^{n-1}+\ldots+c_{1}x+c_{0}\in R[x,\theta_{i}]/(x^{n}-1)$ which are left multiple of one element $G\in R[x,\theta_{i}]/(x^{n}-1)$(the generator polynomial). The multiplication is defined by the basic rule $(ax^{i})(bx^{j})=a\theta^{i}(b)x^{i+j}$, but this multiplication is not commutative.

\textbf{Lemma 4.1} ([10]) A linear code of length $n$ over $\mathbb{F}_{q}$ is a skew cyclic code if and only if it is a left $\mathbb{F}_{q}[x,\theta]$-submodule of  $\mathbb{F}_{q}[x,\theta]/(x^{n}-1)$. Moreover, if $C$ is a left submodule of $\mathbb{F}_{q}[x,\theta]/(x^{n}-1)$, then $C$ is generated by a monic polynomial $g(x)$ which is a right divisor of $x^{n}-1$ in $\mathbb{F}_{q}[x,\theta]$.

\textbf{Theorem 4.1} Let $C$ be a linear code over $R$ of length $n$ and $C=\eta_{1}C_{1}\oplus\eta_{2}C_{2}\oplus\eta_{3}C_{3}$, where $C_{1}, C_{2}$ and $C_{3}$ are codes over $\mathbb{F}_{q}$ of length $n$, then $C$ is a skew cyclic code with respect to the automorphism $\theta_{i}$ if and only if $C_{1}, C_{2}$ and $C_{3}$ are skew cyclic codes over $\mathbb{F}_{q}$ with respect to the automorphism $\theta_{i}$.

\textbf{Proof} Let $(a_{0},a_{1},\ldots,a_{n-1})\in C_{1}, (b_{0},b_{1},\ldots,b_{n-1})\in C_{2}$ and $(c_{0},c_{1},\ldots,c_{n-1})\in C_{3}$. Assume that $r_{i}=\eta_{1}a_{i}+\eta_{2}b_{i}+\eta_{3}c_{i}$ for $i=0,1,\ldots,n-1,$ then the vector $(r_{0},r_{1},\ldots,r_{n-1})\in C$. If $C$ is a skew cyclic code then $(\theta_{i}(r_{n-1}),\theta_{i}(r_{0}),\ldots,\theta_{i}(r_{n-2}))\in C$, note that $\sigma(r)=(\theta_{i}(r_{n-1}),\theta_{i}(r_{0}),\ldots,\theta_{i}(r_{n-2}))=\eta_{1}(a^{p^{i}}_{n-1},a^{p^{i}}_{0},
\ldots,a^{p^{i}}_{n-2})+\eta_{2}(b_{n-1}^{p^{i}},b_{0}^{p^{i}},\ldots,b_{n-2}^{p^{i}})+\eta_{3}(c_{n-1}^{p^{i}},c_{0}^{p^{i}},\\\ldots,\
c_{n-2}^{p^{i}})$.
Hence, $(\theta_{i}(a_{n-1}),\theta_{i}(a_{0}),\ldots,\theta_{i}(a_{n-2}))=(a_{n-1}^{p^{i}},a_{0}^{p^{i}},\ldots,a_{n-2}^{p^{i}})\in C_{1}, (\theta_{i}(b_{n-1}),\\\theta_{i}(b_{0}),\ldots,\theta_{i}(b_{n-2}))\in C_{2}$,
$(\theta_{i}(c_{n-1}),\theta_{i}(c_{0}),\ldots,\theta_{i}(c_{n-2}))\in C_{3}$,
which implies that $C_{1},C_{2},C_{3}$ are skew cyclic codes over $\mathbb{F}_{q}$.

On the other hand, suppose that $C_{1}, C_{2}$ and $C_{3}$ are all skew cyclic codes over $\mathbb{F}_{q}$ and $(r_{0},r_{1},\ldots,r_{n-1})\\\in C$, where $r_{i}=\eta_{1}a_{i}+\eta_{2}b_{i}+\eta_{3}c_{i}$ for $i=0,1,\ldots,n-1$, then $(a_{0},a_{1},\ldots,a_{n-1})\in C_{1}, (b_{0},b_{1},\ldots,b_{n-1})\in C_{2}$ and $(c_{0},c_{1},\ldots,c_{n-1})\in C_{3}$. Note that $\sigma(r)=(\theta_{i}(r_{n-1}),\theta_{i}(r_{0}),\ldots,\\\theta_{i}(r_{n-2}))=\eta_{1}(a^{p^{i}}_{n-1},a^{p^{i}}_{0},
\ldots,a^{p^{i}}_{n-2})+\eta_{2}(b_{n-1}^{p^{i}},b_{0}^{p^{i}},\ldots,b_{n-2}^{p^{i}})+\eta_{3}(c_{n-1}^{p^{i}},c_{0}^{p^{i}},\ldots,c_{n-2}^{p^{i}})
=\\\eta_{1}(\theta_{i}(a_{n-1}),\theta_{i}(a_{0}),\ldots,\theta_{i}(a_{n-2}))+\eta_{2}(\theta_{i}(b_{n-1}),\theta_{i}(b_{0}),\ldots,\theta_{i}(b_{n-2}))+\eta_{3}
(\theta_{i}(c_{n-1}),\theta_{i}(c_{0}),\ldots,\\\theta_{i}(c_{n-2}))\in \eta_{1}C_{1}\oplus\eta_{2}C_{2}\oplus\eta_{3}C_{3}=C$, so $C$ is a skew cyclic code over $R$.

From Theorem 4.1, we can easily prove the following corollary.

\textbf{Corollary 4.1} If $C$ be a skew cyclic code over $R$, then the dual code $C^{\perp}$ is also skew cyclic.

\textbf{Proof} By Theorem 3.2, we have $C^{\perp}=\eta_{1}C_{1}^{\perp}\oplus\eta_{2}C_{2}^{\perp}\oplus\eta_{3}C_{3}^{\perp}$.
According to [3, Corollary 18], we know that the dual code of every skew cyclic code over $\mathbb{F}_{q}$ is also skew cyclic. Hence the dual code $C^{\perp}$ is a skew cyclic code from Theorem 4.1.

\textbf{Definition 4.2} Let $\mathscr{C}$ be a linear code of length $n$ over $\mathbb{F}_{q}$ and $c=(c_{0},c_{1},\ldots,c_{n-1})=(c^{1}|c^{2}|\ldots|c^{l})$ be a codeword in $\mathscr{C}$ divided into $l$ equal parts of length $m$ where $n=ml$. If $\varphi_{l}=(\sigma(c^{1})|\sigma(c^{2})|\ldots|\sigma(c^{l}))\in \mathscr{C}$, where $\varphi$ is the usual cyclic shift of $C$, then the linear code $C$ which is permutation equivalent to $\mathscr{C}$ is called a skew quasi-cyclic code of index $l$.

The next corollary follows from the definition of quasi-cyclic codes.

\textbf{Corollary 4.2} If $C$ is a skew cyclic code of length $n$ over $R$, then $\Phi(C)$ is a skew 3-quasi cyclic code of length $3n$ over $\mathbb{F}_{q}$.

\textbf{Proof} The result follows from the Definition 4.2 and Definition 2.1.

 We are now ready to consider the generator polynomial of a skew cyclic code with length $n$ over $R$.

\textbf{Theorem 4.2} Let $C=\eta_{1}C_{1}\oplus\eta_{2}C_{2}\oplus\eta_{1}C_{3}$ be a skew cyclic code of length $n$ over $R$ and assume that $g_{1}(x), g_{2}(x)$ and $g_{3}(x)$ are generator polynomials of $C_{1}, C_{2}$ and $C_{3}$, respectively, then $C=\langle\eta_{1}g_{1}(x),\eta_{2}g_{2}(x),\eta_{3}g_{3}(x)\rangle$ and $|C|=q^{3n-\sum_{i=1}^{3}deg(g_{i}(x))}$.

\textbf{Proof} Since $C_{1}=\langle g_{1}(x)\rangle, C_{2}=\langle g_{2}(x)\rangle, C_{3}=\langle g_{3}(x)\rangle$, $|C_{i}|=q^{n-deg(g_{i}(x))}, i=1,2,3$, and $C=\eta_{1}C_{1}\oplus\eta_{2}C_{2}\oplus\eta_{3}C_{3}$, then
\begin{eqnarray*}
C&=&\{c(x)=\eta_{1}k_{1}(x)g_{1}(x)+\eta_{2}k_{2}(x)g_{2}(x)+\eta_{3}k_{3}(x)g_{3}(x)\mid\\&&
 k_{1}(x),k_{2}(x),k_{3}(x)\in F_{q}[x,\theta_{i}]\},
\end{eqnarray*}
so, $C\subseteq\langle\eta_{1}g_{1}(x),\eta_{2}g_{2}(x),\eta_{3}g_{3}(x)\rangle$.

Conversely, let us take $\eta_{1}l_{1}(x)g_{1}(x)+\eta_{2}l_{2}(x)g_{2}(x)+\eta_{3}l_{3}(x)g_{3}(x)\in \langle\eta_{1}g_{1}(x), \eta_{2}g_{2}(x),\\\eta_{3}g_{3}(x)\rangle$, where $l_{1}(x), l_{2}(x), l_{3}(x)\in R[x,\theta_{i}]/(x^{n}-1)$, then $\eta_{1}l_{1}(x)=\eta_{1}k_{1}(x), \eta_{2}l_{2}(x)=\eta_{2}k_{2}(x), \eta_{3}l_{3}(x)=\eta_{1}k_{3}(x)$ for some $k_{1}(x), k_{2}(x), k_{3}(x)\in \mathbb{F}_{q}[x,\theta_{i}].$ Hence $\langle\eta_{1}g_{1}(x),\eta_{2}g_{2}(x),\\\eta_{3}g_{3}(x)\rangle\subseteq C.$
Therefore, $C=\langle\eta_{1}g_{1}(x),\eta_{2}g_{2}(x),\eta_{3}g_{3}(x)\rangle$. Since $|C|=|C_{1}|| C_{2}||C_{3}|$, then we have $|C|=q^{3n-\sum_{i=1}^{3}deg(g_{i}(x))}$.

\textbf{Theorem 4.3} Let $C_{1}, C_{2}$ and $C_{3}$ be skew cyclic codes over $\mathbb{F}_{q}$ and $g_{1}, g_{2}, g_{3}$ be the monic generator polynomials of $C_{1}, C_{2}$ and $C_{3}$, respectively. Suppose that $C=\eta_{1}C_{1}\oplus\eta_{2}C_{2}\oplus\eta_{3}C_{3}$, then there exists a unique polynomial $g(x)\in R[x,\theta_{i}]$ such that $C=\langle g(x)\rangle$ and $g(x)$ is a right divisor of $x^{n}-1$, where $g(x)=\eta_{1}g_{1}(x)+\eta_{2}g_{2}(x)+\eta_{3}g_{3}(x)$.

\textbf{Proof} Let $g(x)=\eta_{1}g_{1}(x)+\eta_{2}g_{2}(x)+\eta_{3}g_{3}(x)$, then it is easy to verify that $\langle g(x)\rangle\subseteq C$. On the other hand $\eta_{1}g_{1}(x)=\eta_{1}g(x), \eta_{2}g_{2}(x)=\eta_{2}g(x), \eta_{3}g_{3}(x)=\eta_{3}g(x)$, which implies that $C\subseteq \langle g(x)\rangle$. Thus $C= \langle g(x)\rangle$.

Since $g_{1}(x), g_{2}(x), g_{3}(x)$ are monic right divisors of $x^{n}-1$ in $\mathbb{F}_{q}[x,\theta_{i}]$, then there are $h_{1}(x), h_{2}(x), h_{3}(x)$ in $\mathbb{F}_{q}[x,\theta_{i}]/(x^{n}-1)$ such that $x^{n}-1=h_{1}(x)g_{1}(x)=h_{2}(x)g_{2}(x)=h_{3}(x)g_{3}(x)$. Thus
\begin{eqnarray*}
[\eta_{1}h_{1}(x)+\eta_{2}h_{2}(x)+\eta_{3}h_{3}(x)]g(x)
&=&[\eta_{1}h_{1}(x)+\eta_{2}h_{2}(x)+\eta_{3}h_{3}(x)]\cdot\\
&&[\eta_{1}g_{1}(x)+\eta_{2}g_{2}(x)+\eta_{3}g_{3}(x)]\\
&=&[\eta_{1}h_{1}(x)g_{1}(x)+\eta_{2}h_{2}(x)g_{2}(x)+\eta_{3}h_{3}(x)g_{3}(x)]\\
&=&[\eta_{1}(x^{n}-1)+\eta_{2}(x^{n}-1)+\eta_{3}(x^{n}-1)]\\
&=&x^{n}-1.
\end{eqnarray*}
Hence $g(x)$ is a right divisor of $x^{n}-1$.

The following corollary follows easily.

\textbf{Corollary 4.3} Every left submodule of $R[x,\theta_{i}]/(x^{n}-1)$ is principally generated.

Let $g(x)=\sum_{i=0}^{r}g_{i}x^{i}$ and $h(x)=\sum_{i=0}^{n-r}h_{i}x^{i}$ be polynomials in $\mathbb{F}_{q}[x,\theta_{i}]$ such that $x^{n}-1=h(x)g(x)$ and $C$ be the skew cyclic code generated by $g(x)$ in $\mathbb{F}_{q}[x,\theta_{i}]/(x^{n}-1)$. Then the dual code of $C$ is a skew cyclic code generated by the polynomial
 $\widetilde{h}(x)=h_{n-r}+\theta_{i}(h_{n-r-1})x+\ldots+\theta_{i}^{n-r}(h_{0})x^{n-r}$([3] Corollary 18).
 
\textbf{Corollary 4.4} Let $C_{1}, C_{2}, C_{3}$ be skew cyclic codes over $\mathbb{F}_{q}$ and $g_{1}, g_{2}, g_{3}$ be their generator polynomials such that $x^{n}-1=h_{1}g_{1}, x^{n}-1=h_{2}g_{2}, x^{n}-1=h_{3}g_{3}$ in $\mathbb{F}_{q}[x,\theta_{i}]$. If $C=\eta_{1}C_{1}\oplus\eta_{2}C_{2}\oplus\eta_{3}C_{3}$, then $C^{\bot}=\langle h(x)\rangle$ where $h(x)=\eta_{1}\widetilde{h_{1}}(x)+\eta_{2}\widetilde{h_{2}}(x)+\eta_{3}\widetilde{h_{3}}(x)$ and $|C^{\perp}|=q^{\sum_{i=1}^{3}deg(g_{i}(x))}$.

\textbf{Proof} By Theorem 3.2, we have $C^{\perp}=\eta_{1}C_{1}^{\perp}\oplus\eta_{2}C_{2}^{\perp}\oplus\eta_{3}C_{3}^{\perp}$. Since $C_{1}^{\perp}=\langle\widetilde{h_{1}}(x)\rangle, C_{2}^{\perp}=\langle\widetilde{h_{2}}(x)\rangle$, and $C_{3}^{\perp}=\langle\widetilde{h_{3}}(x)\rangle$, we conclude by Theorem 4.3 that $C^{\perp}=\langle h(x)\rangle$.

In the following section, we denote the order of $\theta_{i}$ is $t_{i}=\frac{m}{i}$ for some positive integer and $(n,t_{i})=1$.

\textbf{Lemma 4.2} ([7, Lemma 2]) Let $g(x)\in \mathbb{F}_{q}[x,\theta_{i}]$ be a monic right divisor of $x^{n}-1$. If $(n,t_{i})=1$, then $g(x)\in \mathbb{F}_{p^{i}}[x]$.

The proof of Theorem 4.4 is similar to that of [7, Theorem 6], so we omit the proof here.

\textbf{Theorem 4.4} Let $g(x)\in \mathbb{F}_{q}[x,\theta_{i}]$ be a monic right divisor of $x^{n}-1$ and $C=\langle g(x)\rangle$. If $(n,q)=1$ and $(n,t_{i})=1$, then there exists an idempotent polynomial $e(x)\in \mathbb{F}_{q}[x,\theta_{i}]/(x^{n}-1)$ such $C=\langle e(x)\rangle$.

From Theorem 4.3 and Theorem 4.4, we have the following corollary.

\textbf{Corollary 4.5} Let $C=\eta_{1}C_{1}\oplus\eta_{2}C_{2}\oplus\eta_{3}C_{3}$ be a skew cyclic code of length $n$ over $R$ and $(n,q)=1$, $(n,t_{i})=1$, then $C_{i}$ has the idempotent generator $e_{i}(x), i=1,2,3.$ Moreover, $e(x)=\eta_{1}e_{1}(x)\oplus\eta_{2}e_{2}(x)\oplus\eta_{3}e_{3}(x)$ is an idempotent generator of $C$, i.e $C=\langle e(x)\rangle.$

Gao in [6] showed that a skew cyclic code is equivalent to a cyclic code of length $n$ over $\mathbb{F}_{p}+v\mathbb{F}_{p}$ with some condition, and gave the enumeration of distinct skew cyclic codes of length $n$ over $\mathbb{F}_{p}+v\mathbb{F}_{p}$. From [6], we also give the number of the skew cyclic code of arbitrary length $n$ over $R$.

\textbf{Theorem 4.5} Let $(n,t_{i})=1$ and $x^{n}-1=\prod_{i=1}^{r}p_{i}^{s_{i}}(x)$ where $p_{i}(x)\in \mathbb{F}_{q}[x,\theta_{i}]$ is irreducible. Then the number of skew cyclic codes of length $n$ over $R$ is $\prod_{i=1}^{r}(s_{i}+1)^{3}$.

\textbf{Proof} By Lemma 4.2, if $(n,t_{i})=1$, then $p_{i}(x)\in \mathbb{F}_{q}[x]$. Hence the number of skew cyclic codes of length $n$ over $\mathbb{F}_{q}$ is $\prod_{i=1}^{r}(s_{i}+1)$. By the decomposition theorem, then the number of skew cyclic codes of length $n$ over $R$ is $\prod_{i=1}^{r}(s_{i}+1)^{3}$.
\section{Conclusion}

\hspace*{0.6cm}In this article, we investigate skew cyclic codes over $R=\mathbb{F}_{q}+v\mathbb{F}_{q}+v^{2}\mathbb{F}_{q}$, where $q=p^{m}$, $p$ is a odd prim, and $v^{3}=v$. We give the number of skew cyclic codes of length $n$ over $\mathbb{F}_{q}$ and $\mathbb{F}_{q}+v\mathbb{F}_{q}+v^2\mathbb{F}_{q}$ under certain conditions. We also describe the generator polynomials of skew cyclic codes over the field $\mathbb{F}_{q}$ and $\mathbb{F}_{q}+v\mathbb{F}_{q}+v^{2}\mathbb{F}_{q}$ and investigate the structural properties of skew cyclic codes over $R$ by a decomposition theorem and also show their idempotent generators.

\end{CJK}

\end{document}